\documentclass[pra,twocolumn,showpacs]{revtex4}\usepackage{graphicx,amsmath,amssymb, bm, epstopdf, color,subfigure}

\usepackage{hyperref}
\hypersetup{pdfborder = {0 0 0}}

\newcommand{\bp}{\bm\rho}
\newcommand{\sT}{\mathcal{T}}

\newcommand{\avg}[1]{\langle #1 \rangle}

\newcommand{\mL}{\mathcal{L}}
\newcommand{\mG}{\mathcal{G}}
\begin{document}
\title{Computational ghost imaging versus imaging laser radar for 3D imaging}
\author{Nicholas D. Hardy}
\author{Jeffrey H. Shapiro}
\affiliation{Research Laboratory of Electronics, Massachusetts Institute of Technology, Cambridge, Massachusetts 02139, USA}

\date{\today}

\begin{abstract}
Ghost imaging has been receiving increasing interest for possible use as a remote-sensing system. There has been little comparison, however, between ghost imaging and the imaging laser radars with which it would be competing. Toward that end, this paper presents a performance comparison between a pulsed, computational ghost imager and a pulsed, floodlight-illumination imaging laser radar.  Both are considered for range-resolving (3D) imaging of a collection of rough-surfaced objects at  standoff ranges in the presence of atmospheric turbulence. Their spatial resolutions and signal-to-noise ratios are evaluated as functions of the system parameters, and these results are used to assess each system's performance trade-offs.  Scenarios in which a reflective ghost-imaging system has advantages over a laser radar are identified.
\end{abstract}


\maketitle

\section{Introduction}
Ghost imaging is an active-imaging technique that uses time-varying structured illumination to image a target without spatially-resolving measurements of the light beam that interacts with the target. Traditionally, a beam splitter is used to create two perfectly-correlated beams, the signal and reference, such that the signal interacts with the target and is then measured by a single-pixel bucket detector, while the reference is directly measured by a spatially-resolving detector \cite{Erkmen2010AOP}. As the illumination pattern is varied, the two measurements are correlated until the spatial structure of the target is determined. Neither measurement alone is sufficient to produce the image; it is their cross-correlation that holds the desired target information, i.e., the ghost image.

The first ghost-imaging experiment relied on entangled photon-pairs---obtained from spontaneous parametric downconversion---for its signal and reference fields, so it was believed that the ghost image was a uniquely quantum feature \cite{Pittman1995}, namely nonlocal two-photon interference. Later, ghost imaging was performed with classical pseudothermal light \cite{Scarcelli2006,Ferri2005}, and a controversy arose as to whether these experiments could be explained by the intensity correlation between classical signal and reference fields or \em only\/\rm\ by nonlocal two-photon interference.  Subsequent Gaussian-state analysis provided a unified treatment of downconverter and pseudothermal ghost imaging, showing that the stronger-than-classical correlation of entangled photons yielded better contrast and near-field resolution \cite{Erkmen2008Unified,Erkmen2009SNR,Erkmen2010AOP}, and that the quantum and semiclassical treatments of the pseudothermal imager gave quantitatively-identical performance predictions.  More recently, the intensity cross-correlation versus nonlocal two-photon interference controversy for understanding pseudothermal ghost imaging has been ended through analysis demonstrating that these two explanations can co-exist \cite{Shapiro2012}. 

Once pseudothermal ghost imaging is considered in the framework of structured-illumination imaging, it becomes possible to dispense with a physically-realized reference field.  In particular, deterministic modulation of a laser beam with a spatial light modulator (SLM) can provide the signal field used for target interrogation, while the on-target intensity pattern needed for the reference field can then be calculated via diffraction theory \cite{Shapiro2008Comp}.   The extension of ghost imaging to this computational framework has opened the door for a variety of applications, including demonstration of ghost imaging with phase-sensitive classical light \cite{Venkatraman2011}, and image reconstruction via compressed sensing, instead of correlation \cite{Katz2009}. Pseudothermal ghost imaging has been experimentally done in reflection both in laboratory \cite{Meyers2008} and remote-sensing scenarios \cite{Zhao2012}.   Moreover, recent work analyzing reflective ghost imaging indicates that computational reflective ghost imaging is feasible for remote sensing \cite{Hardy2011},  and computed reference beams can be generated for all target ranges of interest, so that computational ghost imaging has unlimited depth of focus, unlike its pseudothermal counterpart.  So far, however, there has only been a cursory performance comparison between computational ghost imaging and an imaging laser radar for this application \cite{Hardy2010,Erkmen2012}.

In this paper we extend the analysis in \cite{Hardy2011} of a continuous-wave reflective ghost imager to a pulsed, computational ghost-imaging system that is capable of performing three-dimensional imaging, and we compare its behavior to that of a pulsed, floodlight-illumination, imaging laser radar.  We present results for their spatial resolutions and signal-to-noise ratios (SNRs) when imaging rough-surfaced targets that produce fully-developed laser speckle and  the propagation to and from the targets is through atmospheric turbulence. We also investigate the trade-off between spatial resolution and SNR as a function of detector and entrance-pupil sizes. As reflective ghost-imaging systems are only subject to spatial-resolution loss from turbulence in the source-to-target path \cite{Hardy2011,Cheng2009}, whereas a floodlight-illumination laser radar's spatial resolution is only degraded by turbulence in the target-to-receiver path, we consider arbitrary turbulence distributions on all optical paths.
\begin{figure}[ht]
\centering
\includegraphics[width=3.45in]{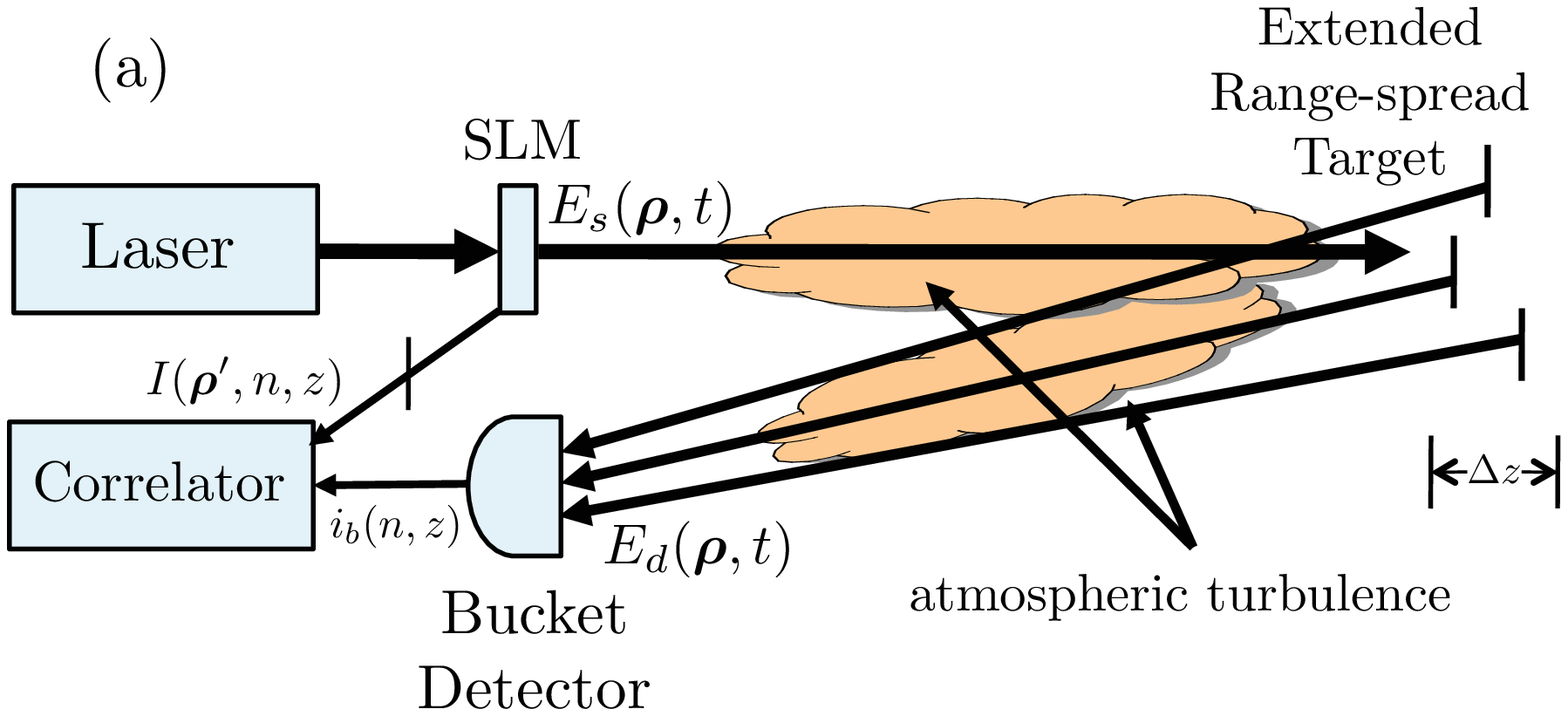}
\\ 
\includegraphics[width=3.4in]{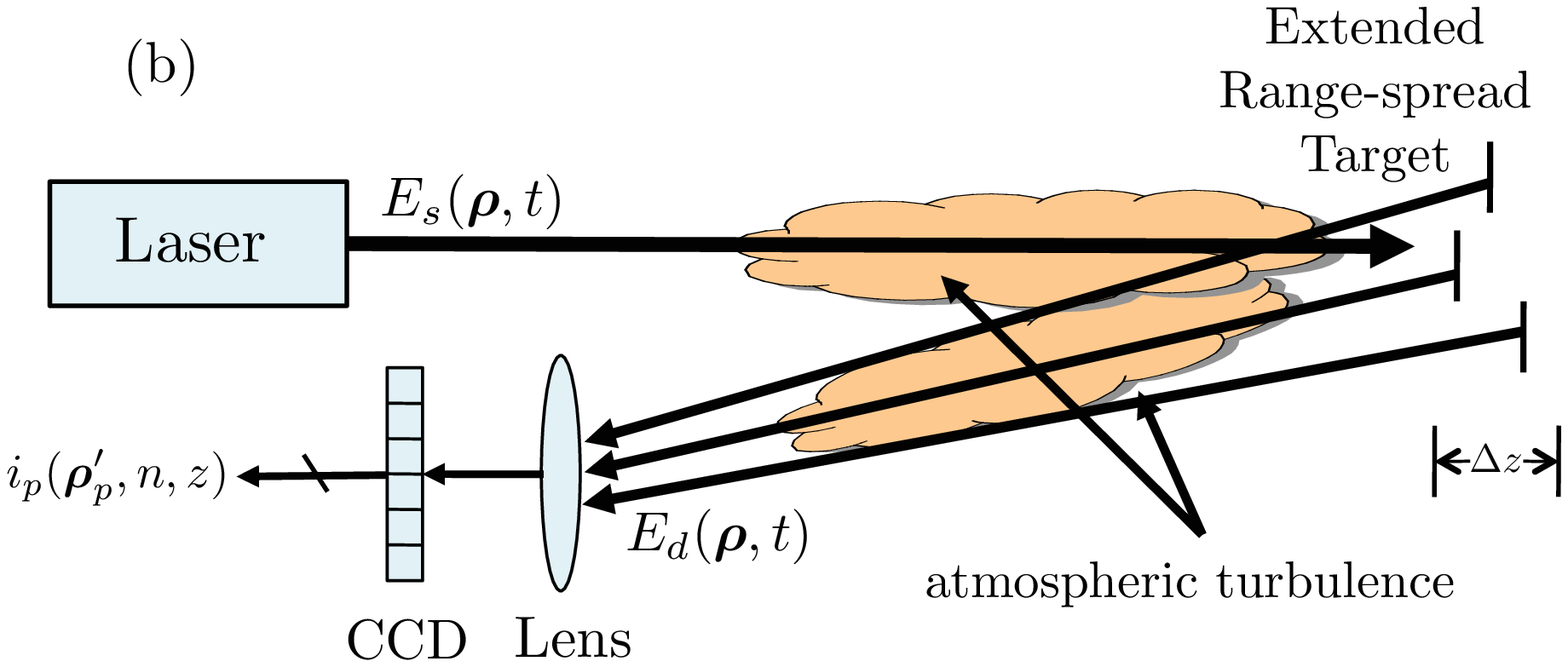}
\label{fig:GI_LR_setup}
\caption{(Color online) (a) Setup for 3D computational ghost imaging in reflection.  Pulsed laser-light undergoes spatial light modulation, propagation through atmospheric turbulence to and from an extended range-spread target, and shot-noise limited bucket detection, producing an output, $i_b(n,z)$, from the range-$z$ return associated with the $n$th transmitted pulse. Diffraction theory is used to calculate $I(\bp',n,z)$, the vacuum-propagation, target-region intensity pattern at transverse coordinate $\bp'$ and range $z$ associated with the $n$th transmitted spatial pattern. (b) Setup for 3D-imaging laser radar. Pulsed laser-light illuminates the target region, in a floodlight manner, through a turbulent optical path. The return light, collected after propagation back to the radar through atmospheric turbulence, is detected by a shot-noise limited CCD array in an image plane, resulting in an output, $i_p(\bp'_p,n,z)$, for the pixel at transverse coordinate $\bp'_p$ and range $z$ from the $n$th transmitted pulse.}
\end{figure}

\section{Source, Propagation, and Targets}\label{sec:Setup}
Figures~1(a) and (b) show the setups that we shall consider for computational ghost imaging and laser radar, respectively.  In this section we highlight the features that are common and different in these systems.  Their performance characteristics will be treated in Secs.~III (ghost imaging) and IV (laser radar).   For both imagers, a pulse-train waveform is emitted by the source and propagates through atmospheric turbulence to an extended, range-spread, rough-surfaced  target.  The light reflected from the target propagates back through atmospheric turbulence to the receiver, where it is photodetected.  For the ghost imager, each pulse has a different spatial pattern, whereas the laser radar uses a constant, floodlight-illumination pattern for all its pulses.  The ghost imager employs a single-pixel bucket detector, and forms its image by cross-correlation with a computed intensity-pattern reference.  The laser radar, however, employs a focusing lens and a CCD array to form an image of the target region.  To make a fair performance comparison between these two systems, we will require that their sources radiate the same average photon number per pulse, $N_T$, towards the target region.  In the subsections that follow, we provide source, propagation, and target details that will serve as the foundation for the performance analyses to come in Secs.~III and IV.

\subsection{Source Characterization}
We will take the SLM output in Fig.~1(a) and the laser output in Fig.~1(b) to be quasimonochromatic, classical scalar waves with center frequency $\omega_0$ (wavelength $\lambda_0 = 2\pi c/\omega_0$, wave number $k_0 = \omega_0/c$) and complex envelope $E_s(\bp, t)$ as a function of transverse coordinate, $\bp$, in the transmitter's exit pupil and time, $t$.   We normalize $E_s(\bp,t)$ so that $|E_s(\bp,t)|^2$ is the photon flux-density emitted from $\bp$ at time $t$.   The pulse-train waveforms for the ghost imager and the laser radar that we will employ are
\begin{equation}
E_s(\bp,t) = \sqrt{N_T} \sum_{n=0}^{N-1} \xi(\bp,n) p(t-n T_s),
\end{equation}
where: $N$ is the number of pulses to be used in forming the image \cite{footnote0}; $\xi(\bp,n)$ is the normalized ($\int\!d\bp\,|\xi(\bp,n)|^2 = 1$) spatial mode of the $n$th pulse; $p(t)$ is a normalized ($\int\!dt\,|p(t)|^2 = 1$) pulse shape that is time-limited to $|t|\le T_p/2$; and $T_s > T_p$ is the pulse-repetition interval.  The pulse duration $T_p$, which could be picoseconds to nanoseconds long, will be taken to be much shorter than the coherence time, $T_c$, of the turbulence, which is typically milliseconds long.  The pulse-repetition interval, $T_s$, will be long enough to preclude second-time-around echoes from the farthest range of interest to masquerade as returns from a closer range \cite{Skolnik}.  Also, $T_s$ will be allowed to exceed $T_c$.

The principal difference between the fields transmitted by the ghost imager and the laser radar lies in their normalized spatial modes, $\xi(\bp,n)$.  For the ghost imager these will be modeled as a collection of independent, identically distributed, zero-mean Gaussian random fields---indexed by the pulse number $n$---that are completely characterized by the following Gaussian Schell-model correlation functions \cite{footnote2}
\begin{eqnarray}
\label{eq:xiCorr1}
\langle \xi(\bp_1,n_1)\xi(\bp_2,n_2)\rangle &=& 0, \\[.1in]
\label{eq:xiCorr2}
\langle \xi^*(\bp_1,n_1)\xi(\bp_2,n_2)\rangle &=& \frac{2}{\pi a_0^2}\,e^{-(|\bp_1|^2+|\bp_2|^2)/a_0^2} \nonumber \\[.1in]
&\times& e^{-|\bp_1-\bp_2|^2/2\rho_0^2}.
\end{eqnarray}
Here, $a_0$ is the source's intensity radius and $\rho_0 \ll a_0$ is its spatial coherence length.  The Gaussian Schell-model is an analytical convenience that captures the essential physics of radiation from a partially-coherent source.  The reader is cautioned, however, that the Gaussian statistics we have assumed will not be valid until sufficient propagation away from the SLM has occurred that the Central Limit Theorem can be applied to the superposition of the many phase-modulated field elements from that modulator's pixels.  Inasmuch as we will be taking propagation from a single SLM pixel to be in its far-field regime, the Gaussian assumption will indeed be applicable for the ghost imager's target illumination.  

The normalized spatial mode for the laser radar will be deterministic, identical for all pulses, and given by the collimated Gaussian beam
\begin{equation}
\xi(\bp,n) = \sqrt{\frac{2}{\pi w_o^2}}\,e^{-|\bp|^2/w_o^2}.
\end{equation}
where $w_o$ is its beam waist.  For a fair comparison with the ghost imager, we shall set the laser radar's beam waist $w_o$ equal to the ghost imager's $\rho_o$.  By doing so, the average far-field intensity pattern produced by the latter's source spatial-mode will exactly match the far-field intensity pattern produced by that of the former.  

\subsection{Propagation through Turbulence}
Propagation of $E_s(\bp,t)$ from the ghost imager or the laser radar to range $z$ in the target region---and propagation of a reflected field $E_r(\bp',t)$ from range $z$ back to those sensors---will be through atmospheric turbulence, whose behavior we will characterize via the extended Huygens-Fresnel principle \cite{Strohbehn1978}.  Labeling the path from the sources to the target region with $S$, and the path leading to the detectors with $D$, we have that
\begin{equation}\label{eq:PropS}
E_z(\bp',t) = \int\! d\bp \, E_s(\bp,t-z/c)h^S_z(\bp',\bp,t),
\end{equation}
and
\begin{equation}\label{eq:PropD}
E_d(\bp,t) = \int\! d\bp \, E_r(\bp',t-z/c)h^D_z(\bp,\bp',t),
\end{equation}
where 
\begin{equation}\label{eq:PropKernelS}
h^S_z(\bp',\bp,t) = \frac{k_0 e^{i k_0(z+|\bp'-\bp|^2/2z)}e^{\psi_S(\bp',\bp,t)}}{i 2 \pi z}
\end{equation}
and
\begin{equation}\label{eq:PropKernelD}
h^D_z(\bp,\bp',t) = \frac{k_0 e^{i k_0(z+|\bp-\bp'|^2/2z)}e^{\psi_D(\bp,\bp',t)}}{i 2 \pi z}
\end{equation}
specify the atmospheric propagation kernels at time $t$ in terms of the Fresnel-diffraction Green's function and the prevailing atmospheric fluctuations, $\psi_S(\bp',\bp,t)$ and $\psi_D(\bp,\bp',t)$, arising from Kolmogorov-spectrum turbulence distributed, in general, non-uniformly along two paths.  The real and imaginary parts, $\chi_S(\bp',\bp,t)$ and $\phi_S(\bp',\bp,t)$, of $\psi_S(\bp',\bp,t)$ are the logamplitude and phase fluctuations imposed on the field arriving $z$\,m downrange at time $t$ and transverse coordinate $\bp'$ from a point source at transverse coordinate $\bp$; there is a corresponding physical interpretation for the real and imaginary parts,  $\chi_D(\bp,\bp',t)$ and $\phi_D(\bp,\bp',t)$, of $\psi_D(\bp,\bp',t)$.  

The range-spread targets of interest will be assumed to lie at ranges between $z_{\rm min}$ and $z_{\rm max} = z_{\rm min} + \Delta z$ from the two imagers under consideration, where $z_{\rm min}$ satisfies the far-field propagation conditions, $k_0a_0\rho_0/2z_{\rm min} \ll 1$ and $k_0w_0^2/2z_{\rm min} \ll 1$, respectively, for the ghost imager's correlation functions, and laser radar's transmitter beams.  As will be explained below---in conjunction with our spatial-resolution analyses for the ghost imager and the laser radar---we will \em not\/\rm\ assume that far-field propagation conditions automatically apply for the ghost imager's source diameter or the laser-radar receiver's entrance pupil diameter.  

We will assume that the $S$ and $D$ paths are sufficiently separated---by virtue of our imagers having different transmitter exit optics and receiver entrance optics---that $\psi_S(\bp',\bp,t)$ and $\psi_D(\bp,\bp;,t)$ will be statistically independent.  It then turns out that turbulence will only enter in our spatial-resolution calculations through the mutual coherence functions of $\exp[\psi_S(\bp',\bp,t)]$ and $\exp[\psi_D(\bp,\bp',t)]$.  Given our assumption of Kolmogorov spectrum turbulence, the mutual coherence functions we will need are as follows,
\begin{equation}\label{eq:TurbCorrS}
\avg{e^{\psi^*_S(\bp'_1,\bp_1,t)}e^{\psi_S(\bp'_2,\bp_2,t)}} = e^{-D_S(\bp'_1 - \bp'_2, \bp_1 - \bp_2)/2},
\end{equation}
and
\begin{equation}\label{eq:TurbCorrD}
\avg{e^{\psi^*_D(\bp_1,\bp'_1,t)}e^{\psi_D(\bp_2,\bp'_2,t)}} = e^{-D_D(\bp_1 - \bp_2, \bp'_1 - \bp'_2)/2},
\end{equation}
where
\begin{equation}
D_S(\bp',\bp) = 2.91 k_0^2z \int_0^1 \!ds\, C^2_{n,S}(s z)\left|\bp' s +\bp (1-s)\right|^\frac{5}{3},
\label{eq:StructS}
\end{equation}
and
\begin{equation}
D_D(\bp,\bp') = 2.91 k_0^2z \int_0^1 \!ds\, C^2_{n,D}(s z)\left|\bp (1-s) +\bp' s\right|^\frac{5}{3},
\label{eq:StructD}
\end{equation}
with $\{\,C^2_{n,S}(\zeta) : 0 \le \zeta \le z\,\}$ and $\{\,C^2_{n,D}(\zeta) : 0 \le \zeta \le z\,\}$ being the turbulence-strength profiles on the $S$ and $D$ paths from the imagers' location $\zeta=0$ to the target region at range $z$.  In order to obtain closed-form results, however, we shall replace these 5/3-law mutual coherence functions with their square-law approximations, i.e., we will use Eqs.~\eqref{eq:TurbCorrS} and \eqref{eq:TurbCorrD} with
\begin{equation}
D_S(\bp',\bp) =\frac{|\bp'|^2W'_S +\bp'\!\cdot\bp (8/3-W'_S-W_S) + |\bp|^2 W_S}{\rho_S^2},
\label{eq:SquareLawS}
\end{equation}
and
\begin{equation}
D_D(\bp,\bp') = \frac{|\bp|^2W_D +\bp\!\cdot\bp' (8/3-W_D-W'_D) + |\bp'|^2 W'_D}{\rho_D^2}.
\label{eq:SquareLawD}
\end{equation}
In these equations, $\{W_m,W_m'\}$, for $m=S,D$, are path-weighting terms, given by
\begin{eqnarray}
W_m &=& \frac{8}{3}\int_0^1\! ds \, (1-s)^2 \mathcal{C}_{n,m}^2(s z)\\[.1in]
W_m' &=& \frac{8}{3}\int_0^1\! ds \, s^2 \mathcal{C}_{n,m}^2(s z)
\end{eqnarray}
in terms of the normalized turbulence-strength profiles,
\begin{equation}
\mathcal{C}_{n,m}^2(\zeta) = \frac{C_{n,m}^2(\zeta)}{\int_0^1\! ds \,C_{n,m}^2(sz)}, \mbox{ for $m = S,D$,}\end{equation}
and $\{\rho_{S},\rho_{D}\}$ are the spherical-wave turbulence coherence lengths for the $S$ and $D$ paths for uniform turbulence distributions with the same integrated strength as the actual distribution,
\begin{equation}
\label{eq:TCL}
\rho_{m} = \left(1.09k_0^2 z \int_0^1 \! ds \, C_{n,m}^2(s z)\right)^{-3/5}, \mbox{ for $m=S,D$.}
\end{equation}

In our treatment of ghost imaging we will assume that $\rho_0 \ll \rho_S/\sqrt{W_S}$, as will typically be the case, but we will allow $\rho_S/\sqrt{W_S}$ to be larger or smaller than $a_0$.  For the laser radar, we have $w_0 = \rho_0 $, so we can assume $w_0\ll \rho_S/\sqrt{W_S}$, but we will allow $\rho_D/\sqrt{W_D}$ to be larger or smaller than $r_\ell$, which is the radar-receiver's entrance pupil radius. 

\subsection{Target Reflection}
Spatially-coherent, quasimonochromatic light reflected by an opaque object whose surface is rough on the scale of the illumination's wavelength---as most real-world surfaces are---yields speckles in the resulting far-field intensity profile \cite{Goodman2007}. The speckles are due to the superposition of randomly phase-shifted reflections from surface facets possessing microscopic wavelength-scale height variations with correlation lengths on the order of a wavelength.   Averaged over the speckle behavior, the rough surface is a quasi-Lambertian reflector that sends light back into the hemisphere with some average intensity-reflection coefficient.  For both our computational ghost imager and our laser radar, it is this intensity-reflection coefficient, as a function of transverse coordinate and range, that is to be imaged.   In both cases, the speckle fluctuations present a significant impairment to high-quality image formation.  Hence we need to incorporate a statistical model for speckle---relating the field illuminating an extended range-spread target to the field returned to our sensors---into our analysis.  

As suggested in Fig.~1, we will presume the target to comprise a set of $K$ quasi-planar, rough-surfaced objects that are located at discrete ranges, $\{\,z_k : 1\le k \le K\,\}$, obeying $z_{\rm min} \le z_1 < z_2 < \cdots < z_K\le z_{\rm max}$.   Moreover, these objects will be taken to have spatial extents, in transverse dimensions, in excess of our sensors' spatial-resolution capabilities.  Then, the field returned to either the ghost imager or the laser radar is given by
\begin{equation}
E_d(\bp,t) = \sum_{k=1}^K\int\!d\bp'\,T_{z_k}(\bp')E_{z_k}(\bp',t-z_k/c)h_{z_k}^D(\bp,\bp',t),
\label{targetreturn}
\end{equation}
where $E_{z_k}(\bp',t)$, the target illumination at transverse coordinate $\bp'$, range $z_k$, and time $t$, is obtained from Eq.~\eqref{eq:PropS} with $z=z_k$, and $T_{z_k}(\bp')$ is the (random) field-reflection coefficient for the target at range $z_k$.  Note that  Eq.~\eqref{targetreturn} assumes that the reflector at a particular range does not occlude those that are farther away from the sensors.  

To complete our target-return model, we only need to supply statistics for the $\{T_{z_k}(\bp')\}$.  Following laser radar theory \cite{Shapiro1981}, these will be taken to be statistically independent, zero-mean, complex-valued Gaussian random processes characterized by the autocorrelation function
\begin{align}\label{eq:Scatter}
\avg{T_{z_k}^*(\bp'_1)T_{z_k}(\bp'_2)} = \lambda_0^2 \sT_{z_k}(\bp'_1) \delta(\bp'_1-\bp'_2).
\end{align}
where $\sT_{z_k}(\bp')$ is the average intensity-reflection coefficient for the reflector at range $z_k$.   Strictly speaking, the Gaussian statistics cannot apply very close to the target, but, because of the quasi-Lambertian nature of rough-surface reflections, the Central Limit Theorem will apply to $E_d(\bp,t)$, so our assuming that the $\{T_{z_k}(\bp')\}$ have Gaussian statistics is indeed warranted.  

\section{Computational Ghost Imager}
\subsection{Image Formation}
The computational ghost image for range $z$ is formed by cross-correlating the computed intensity patterns, $\{\,I(\bp',n,z) : 0 \le n \le N-1\,\}$, that would be produced from vacuum propagation of the $N$ transmitted spatial patterns $\{\,\xi(\bp,n): 0\le n \le N-1\,\}$, with the corresponding sequence of bucket-detector outputs that have been matched filtered for range $z$, which we denote $\{\,i_b(n,z): 0 \le n \le N-1\,\}$.  The $\{I(\bp',n,z)\}$ are given by
\begin{equation}
I(\bp',n,z) = \left|\int\!d\bp\,\xi(\bp,n)\frac{k_0 e^{i k_0|\bp'-\bp|^2/2z}}{i2\pi z}\right|^2.
\end{equation}
For convenience we have omitted leading constants that would appear if we used a physical reference; this scaling does not affect our analysis. The matched-filtered photocurrents $\{i_b(n,z)\}$ satisfy
\begin{eqnarray}
i_b(n,z) &=& \int\!d\tau\,g(nT_s + 2z/c -\tau) \nonumber \\[.1in]
&\times& \left[q\eta\int\!d\bp\,\mathcal{A}^2_b(\bp)|E_d(\bp,\tau)|^2 + \Delta i_b(\tau)\right],
\end{eqnarray}
where: $g(t) \equiv |p(-t)|^2$ is the matched filter (causality ignored) for the transmitted pulse's intensity; $q$ is the electron charge; $\eta$ is the bucket detector's quantum efficiency; $\mathcal{A}_b(\bp) = e^{-|\bp|^2/r_b^2}$ is the field-transmission pupil function for the bucket detector's photosensitive region \cite{footnote3}; and $\Delta i_b(t)$ is the photocurrent shot noise, which we have assumed to be entirely due to the target return.  (For a treatment of ghost imaging that includes background light and its associated shot noise see \cite{Erkmen2012}.)  

The $N$-pulse ghost image for range $z$ is 
\begin{equation}
\mG_N(\bp',z) = \frac{1}{N-1}\sum_{n=0}^{N-1}\tilde{i}_b(n,z)\Delta I(\bp',n,z),
\label{eq:imgGI}
\end{equation}
where, anticipating the need to approximate the {\sc dc}-block used in continuous-wave pseudothermal ghost imaging to obtain a high-contrast image \cite{Scarcelli2006}, we have chosen to cross-correlate 
\begin{equation}
\tilde{i}_b(n,z) \equiv i_b(n,z) - \frac{1}{N}\sum_{n=0}^{N-1}i_b(n,z)
\end{equation}
with 
\begin{equation}
\Delta I(\bp',n,z) \equiv I(\bp',n,z) - \avg{I(\bp',n,z)},
\end{equation}
for which the ensemble average needed in the second term can be computed because we know the statistics of the spatial-pattern sequence $\{\xi(\bp,n)\}$ being applied to the SLM.  

\subsection{Spatial Resolution}\label{sec:GIres}
The computational ghost imager's spatial resolution is found from the ensemble average of Eq.~\eqref{eq:imgGI}.  We will focus our attention on the ranges at which there are target components, i.e., $z\in \{\,z_k : 1\le k \le K\,\}$.  Furthermore, we shall assume that the ghost imager's pulse duration is short enough to resolve all these ranges; for our $p(t)$ this is guaranteed if $cT_p/2 < \min_{1\le k \le K-1}(z_{k+1}-z_k)$.    Thus we need to evaluate $\avg{\mG_N(\bp',z_k)}$, which simplifies to 
\begin{eqnarray}
\avg{\mG_N(\bp',z_k)} &=& \avg{i_b(n,z_k)I(\bp',n,z_k)}  \nonumber \\[.1in]
&-& \avg{i_b(n,z_k)}\avg{I(\bp',n,z_k)}, 
\label{AvGhostImage}
\end{eqnarray}
for $1\le k\le K$, where we have employed the $\{\xi(\bp,n)\}$ being a sequence of statistically independent, identically distributed, spatial patterns.  

The mean values that appear in the preceding expression are easily computed.  For the computed reference we have
\begin{eqnarray}
\avg{I(\bp',n,z_k)} &=& \int\!d\bp_1\!\int\!d\bp_2\,\avg{\xi^*(\bp_1,n)\xi(\bp_2,n)}   \nonumber \\[.1in]
&\times& \frac{k_0^2e^{-ik_0[|\bp_1|^2-|\bp_2|^2 - 2\bp'\cdot(\bp_1-\bp_2)]/2z_k}}{(2\pi z_k)^2} 
\nonumber \\[.1in]
& \approx & \frac{2}{\pi a_{z_k}^2}e^{-2|\bp'|^2/a_{z_k}^2},
\label{intpatt}
\end{eqnarray}
where $a_{z_k} \equiv \lambda_0z_k/\pi \rho_0$ is the intensity radius of $|E_{z_k}(\bp',t)|^2$, and the approximation follows from our assumptions that $\rho_0 \ll a_0$ and $k_0a_0\rho_0/2z_k \ll 1$.  For the bucket detector we find
\begin{eqnarray}
\hspace*{-.1in}\avg{i_b(n,z_k)} &=&  q\eta N_T \int\!d\tau\,g(nT_s+2z_k-\tau) \nonumber \\[.1in]
&\times& \int\!d\bp_b\,\mathcal{A}_b^2(\bp_b)\avg{|E_d(\bp_b,\tau)|^2} \\[.1in]
&\approx& \frac{2q\eta N_TA_b}{\pi a_{z_k}^2T_p'z_k^2}\int\!d\bp'\,\mathcal{T}_{z_k}(\bp')e^{-2|\bp'|^2/a_{z_k}^2},
\label{GIfov}
\end{eqnarray}
where $\rho_0 \ll \min(a_0,\rho_S/\sqrt{W_S})$ justifies the approximation, $A_b \equiv \int \! d\bp_b \, \mathcal{A}_b^2(\bp_b) = \pi r_b^2/2$ is the area of the bucket detector's photosensitive region, and $T_p' \equiv 1/\int\!dt\,|p(t)|^4$ is the effective duration of the matched filter's response to $|p(t)|^2$.  The exponential term appearing in Eq.~\eqref{GIfov} comes from the average intensity pattern in the $z_k$ plane, as seen in Eq.~\eqref{intpatt}.  Going forward, we shall assume that the target lies well within the center of this average intensity pattern, which reduces Eq.~\eqref{intpatt} to
\begin{equation}
\avg{I(\bp',n,z_k)} \approx \frac{2}{\pi a_{z_k}^2},
\end{equation}
on the range-$z_k$ target, and Eq. \eqref{GIfov} to
\begin{equation}
\avg{i_b(n,z_k)} \approx \frac{2q\eta N_TA_b}{\pi a_{z_k}^2T_p'z_k^2}\int\!d\bp\,\mathcal{T}_{z_k}(\bp).
\end{equation}

To complete our evaluation of the average range-$z_k$ ghost image, $\avg{\mG_N(\bp',z_k)}$, 
we need derive an expression for $\avg{i_b(n,z_k)I(\bp', n,z_k)}$.  Its derivation is more involved than what sufficed above for $\avg{I(\bp',n,z)}$ and $\avg{i_b(n,z_k)}$, but, because it parallels similar analysis from \cite{Hardy2011}, we shall merely describe the procedure and present the result.   First, we backpropagate the bucket-detector measurements to the $z=0$ source plane.  Next, we exploit the statistical independence of the sequence of spatial patterns, the turbulence present on the source-to-target and target-to-receiver paths, and the target's field-reflection coefficient.  We then employ our far-field propagation conditions, the correlation function of the target's field-reflection coefficient, and the mutual coherence functions for the turbulence fluctuations incurred on the two paths. Finally, to evaluate a fourth-order moment of the spatial patterns $\xi(\bp,n)$ we employ Gaussian moment-factoring, and thus express the fourth-order moment in terms of second-order moments given in Eqs.~\eqref{eq:xiCorr1} and \eqref{eq:xiCorr2}. The average ghost image term in Eq.~\eqref{AvGhostImage} is then found to be
\begin{equation}
\avg{\mG_N(\bp',z_k)} = \frac{4 q \eta N_T A_b\rho_{z_k}^2}{\pi a^4_{z_k} T_p' z_k^2}\int\!d\bp\,\mathcal{T}_{z_k}(\bp)\frac{e^{-|\bp'-\bp|^2/\alpha\rho_{z_k}^2}}{\pi \alpha\rho_{z_k}^2}\!,
\label{eq:ghostimage}
\end{equation}
with $\rho_{z_k} \equiv \lambda_0z_k/\pi a_0$ being the range-$z_k$ coherence length of the transmitted spatial patterns $\{\xi(\bp,n)\}$, and $\alpha \equiv 1+ a_0^2W_S/2\rho_S^2$ the resolution-degradation factor imposed by the turbulence present in the source-to-range-$z_k$ path. 

Equation~\eqref{eq:ghostimage} shows that the average computational ghost image for the target component at range $z_k$ is $\sT_{z_k}(\bp)$ convolved with a Gaussian point-spread function (PSF) of width $\sqrt{\alpha}\,\rho_{z_k}$.  In the absence of turbulence $\alpha= 1$, so that we get the spatial resolution previously found for single-range, continuous-wave operation in \cite{Shapiro2008Comp}, viz., the speckle coherence length.  In the presence of turbulence in the source-to-target path we have $\alpha > 1$, and our result coincides with that for single-range, continuous-wave operation through turbulence given in \cite{Hardy2011}.  The key points to be gleaned from Eq.~\eqref{eq:ghostimage} are:  (1) turbulence in the source-to-target path does not significantly degrade spatial resolution until its spherical-wave coherence length in the source plane, $\rho_S/\sqrt{W_S}$, becomes smaller than the source's intensity radius, $a_0$; (2) the spatial incoherence of the rough-surfaced target leads to turbulence in the target-to-receiver path having no effect on the average ghost image; (3) computational ghost imaging does not suffer from the turbulence that a pseudothermal configuration would were its propagation path through turbulent air; and (4) our correlating $\tilde{i}_b(n,z)$ with $\Delta I(\bp',n,z)$ has eliminated the background term from $\avg{\mG_N(\bp',z)}$---which would have limited image contrast had we instead correlated $i_b(n,z)$ with $I(\bp',n,z)$---resulting in a high-contrast image.  

A final point to be made here concerns depth of focus and coherence propagation. Our far-field assumption for the ghost imager, $k_0a_0 \rho_0/2z_k \ll 1$, is an analytic tool that simplifies the propagation of the \emph{ensemble-averaged} correlation function in Eq. \eqref{eq:xiCorr2} from the source to range $z_k$.  However, it does not imply that \emph{specific} field patterns can be calculated with a far-field approximation, i.e., a spatial Fourier transform, so that the intensity patterns at different ranges differ only by coordinate and amplitude scaling. Because $\rho_0 \ll a_0$, there is a significant region wherein $k_0a_0 \rho_0/2z_k \ll 1$, but we are in the near field of the source's intensity diameter, viz., $k_0a_0^2/2z_k \gg 1$. In this regime, it is important that the reference used to form the range-$z_k$ ghost image be computed specifically for that range.  This is because computational ghost-image spatial resolution, in the absence of turbulence, is known to degrade by a factor of $\sqrt{1+ (\delta z/z_k)^2(k_0a_0^2/4z_k)^2}$, where $\delta z \equiv z-z_k$,  when the reference used was computed for range $z$ instead of range $z_k$ \cite{Shapiro2008Comp}.  So, deep in the near field a small range mismatch, $|\delta z|/z_k \ll 1$,  between the reference and the target can substantially degrade the spatial resolution.  The same degradation applies to a pseudothermal ghost imager, in which a physical reference arm is used, hence in that case a \em different\/\rm\ reference measurement must be made for each target range of interest if the range spread exceeds that imager's depth of focus.  

\subsection{Signal-to-Noise Ratio}\label{sec:GI_SNR}
We will now evaluate the ghost imager's signal-to-noise ratio, defined to be ratio of $\mG_N(\bp',z_k)$'s squared mean to its variance,
\begin{equation}
\text{SNR}_{\mG_N}(\bp',z_k) \equiv \frac{\avg{\mG_N(\bp',z_k)}^2}{\avg{\mG_N^2(\bp',z_k)} - \avg{\mG_N(\bp',z_k)}^2}.
\end{equation}
To simplify the derivation we make several additional assumptions. First, that the ghost imager resolves all significant detail in the target, allowing us to use $\int \! d\bp \,\sT_{z_k}(\bp)e^{-|\bp'-\bp|^2/\alpha\rho_{z_k}^2}\approx \pi \alpha\rho_{z_k}^2\sT_{z_k}(\bp')$. With this assumption, the square of Eq.~\eqref{eq:ghostimage} becomes
\begin{equation}
\label{eq:GI_mean_squared}
\avg{\mG_N(\bp',z_k)}^2 = \left[\frac{4q \eta N_T A_b\rho_{z_k}^2}{\pi a^4_{z_k} T_p' z_k^2}\sT_{z_k}(\bp')\right]^2.
\end{equation}
This leaves the tedious task of finding $\avg{\mG_N^2(\bp',z_{k})}$,  which involves:  eighth-order and sixth-order moments of the $\{\xi(\bp,n)\}$; propagation through turbulence to the target element at range $z_k$, including a fourth-moment evaluation for the incurred turbulence fluctuations;  target reflection, including a fourth-moment evaluation for the field-reflection coefficient; propagation back to the receiver, requiring another turbulence-fluctuation fourth moment evaluation; and photodetection, with is accompanying shot noise and matched filtering.  The statistical independence of the sequence of spatial patterns, the turbulence fluctuations on each path, and the target's field-reflection coefficient is a considerable help in completing this evaluation.  So too are the Gaussian distributions of the $\{\xi(\bp,n)\}$ and $T_{z_k}(\bp')$, which permit high-order moments to be found from the second moments we have presented earlier via Gaussian moment-factoring.  

The most vexing difficulty in the calculation turns out to come from the fourth moments of the turbulence fluctuations.  Here we shall follow the lead provided in \cite{Hardy2011}, where the corresponding SNR for single-range, continuous-wave operation was derived.  Our work, however, will differ from that in \cite{Hardy2011} in that it will account for pulsed operation in which a single pulse's duration, $T_p$ is much shorter than the atmospheric coherence time, but $NT_s$, the time duration of the $N$-pulse sequence used to form the ghost image, will be taken to span $1 \le N_c \le N$ atmospheric coherence times.  

Following \cite{Hardy2011} we take the logamplitude fluctuations to be Gaussian distributed, and assume that the turbulence is of weak-to-moderate strength---or sufficiently concentrated near the target---that we can both ignore the coordinate dependence of the turbulence at the transmitting and receiving planes, and assume that the logamplitude coherence length at the range-$z_k$ target is larger than $\rho_{z_k}$. Under these conditions, the relevant turbulence-fourth moments for the source-to-target and target-to-detector paths reduce to $e^{4\sigma_S^2 K_S[(n-n')T_s]}$ and $e^{4\sigma_D^2K_D[(n-n')T_s]}$, respectively, for the correlation between turbulence affecting pulses $n$ and $n'$.  Here, for $m=S,D$, $K_m(\tau)$ is the logamplitude fluctuation's normalized ($K_m(0) = 1$) covariance function, and
\begin{equation}
\sigma_m^2 = 0.562 \, k_0^{7/6} \int_0^{z_k} \!dz\, C_{n,m}^2(z)\left[\frac{z(z_k-z)}{z_k} \right]^{5/6}
\end{equation}
is its Rytov-approximation variance \cite{Osche2002}. To account for temporal averaging of the turbulence, we define
\begin{equation}
\gamma=\frac{1}{N}\sum_{n=0}^{N-1} \left[e^{4\sigma_S^2K_S(n T_s)+4\sigma_D^2K_D(n T_s)} - 1\right].
\label{turbaverage}
\end{equation}
For $N\gg 1$ in the assumed weak-to-moderate turbulence---for which $\sigma_m^2 \le 0.1$---we will have $\gamma=(e^{4\sigma_S^2 +4\sigma_D^2}-1)/N \ll 1$ when the turbulence decorrelates pulse-to-pulse on both paths, so that $K_S(n T_s)=K_D(n T_s)=\delta_{n0}$. On the other hand, when the turbulence is frozen across all $N$ pulses, so that $K_D(n T_s)=K_S(n T_s)=1$ for all $n$ of interest, we get $\gamma$'s other asymptote, $\gamma= e^{4\sigma_S^2 +4\sigma_D^2}-1$. Finally, to simplify the analysis, we will assume the transmitter pulse is flat-topped, viz., $p(t) = 1/\sqrt{T_p}$ for $-T_p/2\le t \le T_p/2$ \cite{footnote4}.

The preceding tools can now be employed to show that
\begin{equation}
\text{SNR}_{\mG_N}(\bp',z_k) = 
 \frac{\displaystyle \sT^2_{z_k}(\bp')}{\displaystyle \Delta^2\mathcal{S} + \Delta^2\mathcal{R} + \Delta^2\mathcal{D} + \Delta^2\mathcal{F}},
\label{SNRgi}
\end{equation}
with the terms that appear in the noise denominator being
\begin{eqnarray}
\Delta^2\mathcal{S} &=& \frac{A'_{z_k} (1+\beta^{-1})}{\pi \rho_{z_k}^2 N} e^{4 (\sigma_S^2+\sigma_D^2)}, \\[.05in]
\Delta^2\mathcal{R} &=&  \sT^2_{z_k}(\bp')  \frac{1+2\gamma(1+\beta)}{1+2 \beta}, \\[.05in]
\Delta^2\mathcal{D} &=& \frac{A_{z_k} a_{z_k}^2 z_k^2}{2 A_b \pi \eta N_T N \rho_{z_k}^4} , \\[.05in]
\Delta^2 \mathcal{F} &=&  \frac{A_{z_k}^2(e^{4(\sigma_S^2+\sigma_D^2)}-1- \gamma)}{\pi^2 N \rho_{z_k}^4},
\end{eqnarray}
where $A_{z_k} \equiv \int\!d\bp\,\sT_{z_k}(\bp)$ and $A'_{z_k} \equiv  \int\!d\bp\,\sT^2_{z_k}(\bp)$ are two measures of the target's area, and $\beta \equiv r_b^2/a_0^2$ measures the bucket detector's area relative to the source's intensity area.  

These noise terms account for the following phenomena.  The $\Delta^2\mathcal{S}$ term is the noise contribution from the source-produced on-target speckle patterns.  That it decreases inversely with the number of pulses, $N$, used to form the ghost image is indicative of the need to use many different illumination patterns to form a ghost image. This noise term grows linearly with the $\propto$$A'_{z_k}/\pi \rho_{z_k}^2$ number of spatial-resolution cells on the target, and is exacerbated by the presence of turbulence, i.e., $\sigma_S^2 + \sigma_D^2 >0$.  

The $\Delta^2R$ term in the noise denominator arises from the random fading that results from the rough-surface target reflection combined the effects of atmospheric turbulence.  When $NT_s$ spans $N_c \gg 1$ turbulence coherence times, we get $\gamma\rightarrow 0$, indicating that the turbulence contribution to this noise term vanishes.  However, because we have not allowed our rough-surfaced target to decorrelate over this measurement interval, its contribution to this noise term can only be reduced by aperture averaging, i.e., by increasing the bucket detector's area so as to capture and average an increasing collection of uncorrelated target speckles.

The $\Delta^2\mathcal{D}$ noise term is the bucket detector's shot-noise contribution, which is inversely proportional to the average number of detected target-return photons. For fixed optics and a given target, this term is only decreased by increasing the average transmitted photon-number per pulse, $N_T$, or the number of pulses, $N$, used to form the ghost image.  
 
The preceding noise terms were present---albeit with somewhat different scaling factors---in the SNR expression for continuous-time ghost imaging \cite{Hardy2011}, but the $\Delta^2\mathcal{F}$ noise-denominator term in Eq.~\eqref{SNRgi} is a heretofore unencountered consequence of time-varying turbulence during the measurement interval. When the turbulence is frozen across all $N$ pulses, this term disappears, but when the turbulence changes during the measurement it creates randomness in the on-target illumination patterns that changes from pulse to pulse. This new randomness is suppressed by increasing the number of pulses used to form the ghost image. To appreciate the impact of this new term, consider the worst-case scenario, in which the turbulence decorrelates pulse-to-pulse, the new term dominates the noise denominator, and $N$ is to be chosen to achieve a desired SNR while still under the sway of this term.  Here we find that the necessary $N$ value is  proportional to the \em square\/\rm\ of the number of spatial-resolution cells on the target.  In contrast,  were there no time variation to the turbulence---so that the fourth noise term vanished---then  $N$ would only need to be proportional to the number of on-target spatial-resolution cells to achieve the desired SNR when the first and/or third noise terms dominate the second.  Because increasing $N$ with $T_s$ fixed increases the image acquisition time, the fourth noise term can have a significant adverse impact on that acquisition time.  

Having described the various noise contributions to the ghost-image SNR, let us conclude this section by highlighting two physically-important asymptotic forms of Eq.~\eqref{SNRgi}.  When the number of pulses increases without bound, the ghost-image SNR reaches a finite limit, called the saturation SNR, which is given by 
\begin{align}\label{eq:GI_SNRsat}
\text{SNR}_{\mG_N,\text{sat}}(\bp',z_k) = 1+ 2\beta.
\end{align}
Note that it is due solely to the time-independent target speckle \cite{footnote5}, and is independent of the target's range and reflectivity, although the $N_T$ and $N$ values needed to reach the saturation regime do depend on $\sT_{z_k}(\bp')$.  This saturation SNR can only be increased by increasing $\beta$, which means either increasing the size of the bucket detector, or decreasing the size of the source, with the latter entailing a degradation of the ghost imager's spatial-resolution capability.  

When the average number of detected target-return photons is so low that shot noise dominates all other fluctuations in the ghost image, we get the shot-noise-limited SNR
\begin{align}\label{eq:GI_SNRform}
\text{SNR}_{\mG_N,\text{shot}}(\bp',z_k) = \frac{2 \pi \eta N_T N A_b\rho_{z_k}^4  \sT^2_{z_k}(\bp') }{A_{z_k} a_{z_k}^2 z_k^2}.
\end{align}
Unlike the case for its saturation signal-to-noise ratio, the ghost imager's shot-noise limited SNR \em does\/\rm\ depend on the target's range and reflectivity.

\section{Laser Radar}\label{sec:LR}

\subsection{Image Formation}
The $N_p$-pixel laser radar image for range $z$ is formed from the pixel-wise outputs $\{\,i_p(\bp'_p,n,z) : 1\le p \le N_p, 0 \le n \le N-1\,\}$ of a CCD---on which the target-reflected light has been focused by a lens---that have been matched-filtered for range $z$. These photocurrents take the form
\begin{eqnarray}\label{eq:measureLR}
\lefteqn{i_p(\bp'_p,n,z) = \int\!d\tau\,g(nT_s + 2z/c -\tau)} \nonumber \\[.05in]
&\times& \left[q\eta\int\!d\bp'_i\,\mathcal{A}_p^2(\bp'_i)|E_p(\bp'_i,\tau)|^2 + \Delta i_p(\tau)\right],
\end{eqnarray}
where: $\mathcal{A}_p(\bp'_i)$ is the real-valued field-transmission pupil function defining the photosensitive region of the $p$th image-plane pixel, which we take to be centered at $\bp'_i = -\bp'_p$ to compensate for image inversion; $E_p(\bp'_i,\tau)$ is the complex envelope of the light impinging on the CCD plane at point $\bp'_i$ and time $\tau$; and $\Delta i_p(\tau)$ is the photocurrent shot noise for pixel $p$.

In practice the lens' focal length will cast a minified image on the CCD, but, for convenience, we will assume it is chosen to realize 1:1 imaging for range $z_\ell$, which is taken to be the center of the range-interval of interest. It follows that 
\begin{eqnarray}
E_p(\bp',t) &=&  \int\! d\bp \, \mathcal{A}_\ell(\bp)E_d(\bp,t)  \nonumber \\[.05in]
&\times & e^{-i k_0|\bp|^2/z_\ell} \,\frac{k_0 e^{i k_0(z_\ell+|\bp' - \bp|^2/2 z_\ell)}}{i 2 \pi z_\ell},\label{eq:PropP}
\end{eqnarray}
where $\mathcal{A}_\ell(\bp)$ is the lens' real-valued field-transmission pupil function, $E_d(\bp,t)$ is the pupil-plane target-return field given by Eq.~\eqref{targetreturn}, and we have neglected an unimportant absolute phase factor as well as the propagation delay within the radar receiver.  To facilitate analytic comparison with the ghost imager, we will take the lens' pupil function to be  $\mathcal{A}_\ell(\bp) \equiv \exp(-|\bp|^2/r_\ell^2)$. The laser radar then produces its range-$z$ image by pixel-wise averaging the photocurrents from the $N$ pulses,
\begin{equation}\label{eq:LR}
\mL_N(\bp_p',z) = \frac{1}{N}\sum_{n=0}^{N-1} i_p(\bp_p',n,z).
\end{equation}

Before turning to our spatial-resolution analysis, there is an important point to make about depth of focus, cf.\ the discussion at the end of Sec~\ref{sec:GIres} about this issue for the ghost imager. When $k_0r_\ell^2/2z \ll 1$ for all target ranges of interest, the laser radar can be focused at infinity with no loss of spatial resolution at any of those ranges.  However, when $k_0r_\ell^2/2z \gg 1$ for target ranges of interest, then the radar receiver must operate within the depth of focus for the range to be imaged in order to prevent loss of spatial resolution.  The rest of our treatment of the laser radar will assume that the far-field condition, $k_0r_\ell^2/2z \ll 1$, holds, so that all target ranges of interest will be in focus, but this need not always be the case in operational scenarios of interest.    

\subsection{Spatial Resolution}
The laser radar's spatial resolution is found from its average image, just as was done for the ghost imager. As we did in Sec.~\ref{sec:GIres}, we will only consider ranges $\{\,z_k : 1\le k \le K\,\}$ that contain target components, and take the pulse duration, $T_p$, to be short enough to resolve them. We then get
\begin{eqnarray}
\avg{\mL_N(\bp_p',z_k)} &=& q\eta \int\!d\tau\,g(nT_s + 2z_k/c -\tau)   \nonumber\\[.05in] 
&\times&
\int\!d\bp'\,\mathcal{A}_p^2(\bp'_i)\avg{|E_p(\bp'_i,\tau)|^2}.
\end{eqnarray}
Backpropagating $E_p(\bp',\tau)$ to the source using Eqs.~\eqref{eq:PropP}, \eqref{targetreturn}, and \eqref{eq:PropS}, evaluating the resulting source and target second-moments that appear, and then performing all integrations but those over the target and image planes, we obtain
\begin{eqnarray}
\avg{\mL_N(\bp_p',z_k)} &=& \frac{2q\eta N_TA_\ell}{\pi z_k^2 w_{z_k}^2 T_p'}  \int\!d\bp'\,\sT_{z_k}(\bp') e^{-2|\bp'|^2/w_{z_k}^2}\nonumber \\[.05in] 
&\times&
\int\!d\bp'_i\,\mathcal{A}_p^2(\bp'_i)\frac{e^{-|\bp'+\bp'_i|^2/\alpha'\rho_{z_k}^{'2}}}{\pi\alpha'{\rho_{z_k}^{'2}}}.
\label{prepixelint}
\end{eqnarray}
In this expression: $w_{z_k} = \lambda_0z_k/\pi w_0$ is the radar beam's intensity radius at range $z_k$, which we will assume is sufficiently large that $\sT_{z_k}(\bp')e^{-2|\bp'|^2/w_{z_k}^2} \approx \sT_{z_k}(\bp')$; $A_\ell \equiv  \int\! d\bp \, \mathcal{A}_\ell^2(\bp) = \pi r_\ell^2/2$ is the lens' effective area; $\rho_{z_k}' = \lambda_0 z_k/\sqrt{2}\,\pi r_\ell$ is the diffraction-limited (no-turbulence) spatial resolution for an image-plane point detector; and $\alpha' \equiv 1+ r_\ell^2W_D/\rho_D^2$ is a  resolution-degradation factor caused by turbulence.   Turbulence begins to impair the laser radar's spatial resolution when its receiver-plane coherence length, $\rho_D/\sqrt{W_D}$, becomes comparable to the receiver's lens radius, $r_\ell$. Hence, when the turbulence on the return path is either sufficiently weak, or concentrated near the target, we get $\alpha'\approx1$ and there is no loss of resolution. However, for sufficiently strong turbulence, or turbulence concentrated near the receiver's lens, significant resolution degradation is incurred.

The computational ghost imager can calculate its reference field for every $\bp'$ of interest, but the laser radar must use pixels of finite size in order to collect any optical power.  To obtain a closed-form result from the pixel integration in Eq.~\eqref{prepixelint} we will use $\mathcal{A}_p(\bp')\equiv e^{-|\bp'+\bp_p'|^2/r_p^2}$ and arrive at our final expression for the laser radar's average image,
\begin{eqnarray}
\hspace*{-.2in}\avg{\mL_N(\bp_p',z)} &=& \frac{2 q \eta N_T A_\ell A_p }{\pi z_k^2 w_{z_k}^2 T'_p} \nonumber \\[.05in]
&\times&  \int \! d\bp' \, \sT_{z_k}(\bp') \frac{e^{-|\bp'-\bp_p'|^2/( \beta'+\alpha')\rho_{z_k}^{'2}}}{\pi (\beta' +\alpha') \rho_{z_k}^{'2}},
\label{eq:LRres}
\end{eqnarray}
where $A_p \equiv \int\! d\bp' \,\mathcal{A}_p^2(\bp') = \pi r_p^2/2$ is the pixel's effective area, and $\beta' \equiv r_p^2/2 \rho_{z_k}^{'2}$. Equation~\eqref{eq:LRres} quantifies the laser radar's loss of spatial resolution---its degradation from $\rho_{z_k}'$, the diffraction-limited, point-detector value---when the lens area exceeds a turbulence coherence area and/or the pixel area exceeds the diffraction-limited spot size.  

\subsection{Signal-to-Noise Ratio}
The laser radar's SNR for the $p$th pixel's range-$z_k$ value is the ratio of  $\mL_N(\bp'_p,z_k)$'s squared mean to its variance, 
\begin{equation}
\text{SNR}_{\mL_N}(\bp'_p,z_k) \equiv \frac{\avg{\mL_N(\bp_p',z)}^2}{\avg{\mL_N^2(\bp_p',z)} - \avg{\mL_N(\bp_p',z)}^2}.
\end{equation}
Our evaluation of this SNR mirrors what we did for the ghost imager. So, as we did there, we shall assume that all significant target detail is resolved by the imager, and that the turbulence is weak enough that we can ignore its fluctuation-terms' coordinate dependence in the planes of the transmitter, target, and receiver.  With the first assumption, Eq.~\eqref{eq:LRres} yields
\begin{equation}
\avg{\mL_N(\bp'_p,z_k)}^2 = \left[\frac{2 q \eta N_T A_\ell A_p }{\pi z_k^2 w_{z_k}^2 T_p} \sT_{z_k}(\bp'_p) \right]^2.
\end{equation}
As was the case for the ghost imager, the primary difficulty encountered in SNR evaluation is finding the image's second moment.  The laser radar's second-moment calculation, however, is substantially simpler than that for the ghost imager in that the laser radar case only requires turbulence and target fourth-moments and the shot noise's second moment.  Consequently, the evaluation proceeds as follows.  

First, we use the statistical independence of the turbulence, target, and shot noise to separate averages involving these three randomness contributors.  Then we employ Gaussian moment-factoring to reduce the target's fourth moment into a sum of products of second moments, and our assumption about the coordinate-independence of the turbulence fluctuations to evaluate that fourth moment as we did for the ghost imager.  Next, we perform the resulting multi-dimensional Fourier transforms of the Gaussian functions that arise from far-field optical propagation in conjunction with the Gaussian pupil functions we have assumed. Finally, under our assumption that the laser radar resolves all significant target detail, we are left with
\begin{align}
\text{SNR}_{\mL_N}(\bp'_p,z_k) =\frac{\sT_{z_k}(\bp'_p)}{\sT_{z_k}(\bp'_p)(\frac{1 + (2+\beta')\gamma}{1+\beta'}) +  \frac{\pi w_{z_k}^2 z_k^2}{2\eta  N_T N  A_\ell A_p}} ,
\label{LRsnr}
\end{align}
where $\gamma$ is the time-averaged turbulence factor from Eq.~\eqref{turbaverage}.
 
The noise-denominator terms in the preceding SNR formula have the following physical interpretations.  The first term is due to the time-independent target speckle, exacerbated, to some degree, by the turbulence-induced scintillation on the transmitter-to-target and target-to-receiver paths.  The second term is due to shot noise.  Thus when $N_TN$---the average number of photons transmitted over all $N$ pulses---is sufficiently high, the SNR reaches a finite maximum value, namely the saturation signal-to-noise ratio, given by
\begin{equation}\label{eq:LR_SNRsat}
\text{SNR}_{\mL_N,\text{sat}}(\bp'_p,z_k) = 1+\beta',
\end{equation}
which depends only on $r_p^2/2\rho_{z_k}^{'2}$, the number of diffraction-limited spots within the pixel area \cite{footnote5}.    Increasing $\beta'$, e.g., by increasing the pixel size, will increase the saturation value, but doing so degrades the radar's spatial resolution; see Eq.~\eqref{eq:LRres}.  Conversely, when the SNR is much lower than its saturation limit, it takes the shot-noise limited form
\begin{align}\label{eq:LR_SNRform}
\text{SNR}_{\mL_N,\text{shot}}(\bp'_p,z_k) = \frac{2\eta  N_T N A_\ell A_p \sT_{z_k}(\bp'_p)}{\pi w_{z_k}^2 z_k^2}.
\end{align}

\section{Performance Comparison}\label{sec:Comp}
Having completed spatial resolution and SNR analyses for both the computational ghost imager and the laser radar, we are ready to compare their capabilities.  Before proceeding, two points deserve note.  The first concerns ensuring that our spatial resolution and SNR comparisons are fair.  Toward that end we will take $w_0 = \rho_0$, so that the laser radar's on-target intensity pattern matches the ghost imager's average on-target intensity pattern.  We will also assume that $r_\ell = a_0$, because: (1) $r_\ell$ and $a_0$ correspond to intensity radii ($r_\ell$ for the the intensity transmission of the radar receiver, and $a_0$ for the intensity transmission of the ghost imager's exit optics); and (2) $r_\ell$ and $a_0$ determine the diffraction-limited spatial resolutions of these two imagers.  In addition, we we will take $N_{T_{\rm GI}}N_{\rm GI} = N_{T_{\rm LR}}N_{\rm LR}$, i.e., the product of the average number of transmitted photons per pulse and the number of pulses employed for the ghost imager (GI) and the laser radar (LR) must be equal, but we will \em not\/\rm\ require $N_{T_{\rm GI}} = N_{T_{\rm LR}}$ and $N_{\rm GI} = N_{\rm LR}$, as originally indicated in Sec.~\ref{sec:Setup}.  Thus we are constraining both systems to use the same average number of transmitted photons to form their images, but we are affording the laser radar the opportunity to exploit its ability to form an image from a single pulse, in contrast to the ghost imager's fundamental requirement for averaging the returns from multiple pulses.

Our second pre-comparison note is cautionary.  Our analysis has employed Gaussian functions for partially-coherent and coherent optical beams, and for photodetector pupil-functions, etc., in order to obtain closed-form expressions for our systems' spatial resolutions and SNRs.  Although the use of Gaussian functions should yield the correct dependence of these performance metrics on system parameters, the constant factors that appear in our results would be different had we numerically evaluated the spatial resolutions and SNRs for circular-pupil optics.  

\subsection{Spatial-Resolution Comparison}
We have shown that the ghost imager and laser radar's average images for the target component at range $z_k$ convolve $\sT_{z_k}(\bp')$ with Gaussian point-spread functions of the form 
$P(\bp') = e^{-|\bp'|^2/r_{\rm res}^2}/\pi r_{\rm res}^2,$ where
\begin{equation}
r_{\rm res} = \left\{\begin{array}{ll}
\sqrt{\alpha}\,\rho_{z_k}, & \mbox{ghost imager} \\[.05in]
\sqrt{\beta' + \alpha'}\,\rho_{z_k}' ,& \mbox{laser radar},\end{array}\right.
\label{rescomp}
\end{equation}
with $\rho_{z_k}$ and $\rho'_{z_k}$ being the diffraction-limited resolutions, $\alpha \ge 1$ and $\alpha' \ge 1$ accounting for turbulence-induced resolution loss, and $\beta' \ge 0$ for resolution lost because of finite pixel size.

The diffraction-limited resolutions are $\rho_{z_k} = \lambda_0 z_k/\pi a_0$ and $\rho'_{z_k} = \lambda_0 z_k/\sqrt{2}\,\pi a_0$, where we have used $r_\ell = a_0$ in the latter, showing that the laser radar's diffraction-limited performance is $\sqrt{2}$-times better than that of the computational ghost imager. The numerical value of this resolution advantage depends on our use of Gaussian functions, but the fact that the laser radar's diffraction-limited spatial resolution is better than that of the computational ghost imager does not, so long as the laser radar's receiver pupil is the same as the ghost imager's transmitter pupil.  That this is so follows from the ghost imager's PSF arising from convolution of the on-target average intensity pattern of its transmitter with the corresponding pattern from the computed reference, while the laser radar's PSF (in the 1:1 imaging setup we have assumed) corresponds to just one of those patterns.   

As noted in Sec.~\ref{sec:LR}, the laser radar must employ finite-sized pixels, and Eq.~\eqref{LRsnr} shows that---other system parameters being held constant---forcing $\beta' \ll 1$ will push the system into its shot-noise limited regime as the pixel area, $A_p$, is decreased.  Thus a prudent compromise might be to size the pixels to satisfy $\beta' = 1$, so that the laser radar's spatial resolution, in the absence of turbulence, only suffers a $\sqrt{2}$-factor increase.  This choice of pixel size, however, wipes out the laser radar's spatial resolution advantage over the computational ghost imager, i.e., the no-turbulence resolution of the former then exactly matches the diffraction-limited resolution of the latter.  

When turbulence is strong enough to dominate the spatial resolution of both imagers, we get
\begin{equation}
r_{\rm res} = \left\{\begin{array}{ll}
\lambda_0 z_k \sqrt{W_S}/\pi\sqrt{2}\,\rho_S, & \mbox{ghost imager} \\[.05in]
\lambda_0 z_k\sqrt{W_D}/\pi\sqrt{2}\,\rho_D, & \mbox{laser radar},\end{array}\right.
\label{turbres}
\end{equation}
where we have continued our use of $r_\ell = a_0$.  For a situation in which $\rho_S/\sqrt{W_S} = \rho_D/\sqrt{W_D}$---such as when the laser radar's receiver is co-located with the ghost-imager's transmitter---we find the turbulence-limited resolution to be equal for both systems. More generally, either system could have more favorable spatial-resolution behavior in the presence of turbulence because different turbulence-strength profiles could exist on the transmitter-to-target and target-to-receiver paths, and the ghost imager's spatial resolution is only sensitive to turbulence on the transmitter-to-target path while the laser radar's spatial resolution is only sensitive to the turbulence on the target-to-receiver path.  

\subsection{Signal-to-Noise Ratio Comparison}
For our SNR comparison we shall consider the two imagers' saturation SNRs and their shot-noise limited SNRs.  Throughout we will assume that any turbulence that might be present has no effect on spatial resolution, so that $\alpha=\alpha'=1$, and that both imagers have sufficient resolution to resolve all significant detail in $\sT_{z_k}(\bp')$.  

Equations~\eqref{eq:GI_SNRsat} and \eqref{eq:LR_SNRsat} specify the saturation SNRs for the ghost imager and the laser radar, and give us
\begin{equation}\label{eq:SNRsatLim}
\frac{\text{SNR}_{\mL_N\text{,sat}}(\bp'_p,z_k)}{\text{SNR}_{\mG_N\text{,sat}}(\bp'_p,z_k)} = 
\frac{1+\beta'}{1+2\beta}.
\end{equation} 
Setting $r_\ell = a_0$, to make the laser radar's receiver pupil be the same size as the ghost imager's transmitter pupil, and $\beta'=1$, so that these systems have the same spatial resolution,  the preceding ratio of their saturation SNRs becomes $2/(1+\beta)$.   Consequently, the computational ghost-imager's saturation SNR will greatly exceed that of the laser radar when $\beta = r_b^2/a_0^2 \gg 1$, i.e., when the bucket detector's area is much larger than the source area.  There is a simple physical explanation for this behavior:  the saturation SNR is due to the time-independent speckle created by reflection from the rough-surfaced target.  These speckles are $\sim$$a_0$ in radius in the receiver's pupil plane, so when $\beta \gg 1$, the bucket detector is averaging over many statistically independent speckles, driving up the ghost imager's saturation SNR.

There is, however, an intrinsic unfairness in the preceding favorable view of the ghost imager's saturation SNR, because by fixing $r_\ell = a_0$ but letting $\beta \gg 1$, we are allowing the bucket detector to have a much larger receiving aperture than the laser radar. Suppose, instead, that we constrain the systems to have the same size receiving aperture, so that $r_b=r_\ell$, but maintain $\beta \gg 1$. In that case the laser radar's diffraction-limited spatial resolution would be better than that of the ghost imager by a factor of $r_b/a_0 \gg 1$ but its SNR would be worse. However, the laser radar could then use $\beta' = 2 r_b^2/a_0^2 \gg 1$ and:  (1) have its no-turbulence spatial resolution match the diffraction limit of the ghost imager; and (2) have its saturation SNR equal  that of the ghost imager.  In this case, the time-independent speckle is being averaged in the laser radar's image plane, because the large pixel comprises a great many independent speckle lobes.  

Finally, if we make all optics the same size, $a_0=r_b=r_\ell$, and choose the pixel size for equal spatial resolutions, then $\beta = \beta' = 1$ and we find $\text{SNR}_{\mL_N\text{,sat}}(\bp'_p,z_k) = 2\,\text{SNR}_{\mG_N\text{,sat}}(\bp'_p,z_k)/3$.

To summarize what we have seen so far concerning saturation-SNR behavior, fair comparisons between the ghost imager and the laser radar indicate that neither system will enjoy a significant advantage in this regard.  There is, however, a fine point to be considered concerning atmospheric turbulence.  We have assumed the relevant turbulence coherence lengths to be long enough that our imagers' spatial resolutions do not lose any resolution arising from propagation over turbulent paths.  Nevertheless, the target-speckle terms in the noise denominators of Eqs.~\eqref{SNRgi} and \eqref{LRsnr} both contain the time-averaged scintillation factor $\gamma$  When a sufficient number of pulses are averaged---as has been assumed in the saturation SNR formulas---we get $\gamma \rightarrow 0$.  For the ghost imager, averaging the returns from a large number of pulses is an intrinsic requirement for image formation, but a laser radar can form its image with a single pulse, which case we get
\begin{equation}
\text{SNR}_{\mL_1}(\bp'_p,z_k) \longrightarrow \frac{1+\beta'}{1+ (2+\beta')\gamma},
\end{equation}
for $N_{T_{\rm LR}}$ sufficiently high.  For $\beta' = 2r_b^2/a_0^2 \gg 1$, this SNR can be substantially \em worse\/\rm\ than the ghost imager's saturation SNR, because $\gamma \ge e^{4\sigma_S^2}$ in this single-pulse case.  The laser radar will then need to employ more than a single pulse to approach its full saturation SNR---and to match that of the computational ghost imager at the same spatial resolution---because of scintillation.

Now let us turn to SNR behavior when neither system's $N_T$ and $N$ values are sufficient to reach SNR saturation by comparing their shot-noise limited performance.  Here, using $r_\ell = a_0$,   Eqs.~\eqref{eq:GI_SNRform} and \eqref{eq:LR_SNRform} give us 
\begin{equation}\label{eq:FormRatio}
\frac{\text{SNR}_{\mL_{N}\text{,shot}}(\bp'_p,z_k)}{\text{SNR}_{\mG_{N}\text{,shot}}(\bp'_p,z_k)}= \frac{A_{z_k} \beta '}{2\pi \rho_{z_k}^2 \beta \mathcal{T}_{z_k}(\bp'_p)}.
\end{equation}
For a fair comparison we again set $\beta = \beta' = 1$, to give the systems the same pupil sizes and spatial resolutions.  The preceding SNR ratio then reduces to $A_{z_k}/2\pi \rho_{z_k}^2\sT_{z_k}(\bp'_p)$.  For a range-$z_k$ target whose intensity-reflection coefficient has limited spatial variation, this number is approximately its number of spatial-resolution cells, making the laser radar's shot-noise limited SNR far superior to that of the computational ghost imager.   Increasing the size of the ghost imager's bucket detector can overcome this SNR disadvantage by making $\beta \gg 1$.  As in our saturation-SNR comparison, however, we should then allow the laser radar to increase its lens size to equal the ghost imager's new $r_b$ value, while operating with $\beta = r_b^2/a_0^2 \gg 1$, to regain its shot-noise limited SNR advantage.  Unlike what we found for the laser radar's saturation SNR, its shot-noise limited SNR is the same at all $N_{T_{\rm LR}}N_{\rm LR}$ values, i.e., Eq.~\eqref{eq:FormRatio} applies even when the laser radar forms its image from a single pulse.  

\section{Conclusions}
Computational ghost imaging is, in many respects, a dual of floodlight-illumination laser radar. The computational ghost imager's system complexity is in its transmitter, whose sequence of SLM patterns creates the structured illumination that provides the imager's spatial resolution.  The laser radar's complexity lies in its CCD receiver, which provides its spatial resolution. Consequently, the size of the ghost imager's transmitter pupil sets its diffraction-limited spatial resolution, whereas the laser radar's no-turbulence spatial resolution is set by the size of its receiver pupil in conjunction with that of its CCD pixels.  Thus only turbulence on the transmitter-to-target path can impair the ghost imager's spatial resolution, while the laser radar's spatial resolution is only impacted by turbulence on the target-to-receiver path. Therefore, ghost imaging and laser radar systems that are designed to have equal spatial resolutions in the absence of turbulence could have significantly different performance in a bistatic configuration, in which their transmitters and and receivers are not co-located, so that significantly different turbulence distributions are encountered on these two paths.  In such situations either one could offer the better spatial-resolution performance, depending on which path had its turbulence concentrated near the resolution-controlling pupil.  Aside from this turbulence issue, our analysis indicates that a fair comparison between the computational ghost imager and the floodlight-illumination laser radar shows them to have equal spatial resolutions, \em except\/\rm\ for the following two caveats:  (1) the laser radar can form its image from a single pulse, making it far better for imaging moving targets; and (2) the computational ghost imager has infinite depth of focus, whereas the laser radar will not for ranges satisfying $k_0r_\ell^2/2z \gg 1$. 

Both the ghost imager and the laser radar have signal-to-noise ratios that are shot-noise limited at low-$N_T$, low-$N$ values and saturate at high-$N_T$, high-$N$ values.  When their optics are sized for equal spatial resolutions, with $N_{T_{\rm GI}}= N_{T_{\rm LR}}$ and $N_{\rm GI} = N_{\rm LR}$, there is little difference in their saturation SNRs.  This contrasts strongly with their shot-noise limited SNR behavior, under these conditions, in which the laser radar outperforms the ghost imager by a factor approximately equal to the number of spatial-resolution cells on the target.  As a result, we can expect that the ghost imager will require significantly more time to  achieve a desired SNR when operating in this regime.  This key disadvantage for correlation-based ghost imaging could be mitigated to some degree, however, by the use of compressed-sensing techniques, which enable many fewer pulses to suffice for ghost-image formation.  Recent work has demonstrated this possibility in table-top ghost imaging done in reflection \cite{Meyers2010}, indicating its likely feasibility for standoff-sensing applications.

What then are the possible advantages of ghost imaging in comparison with laser radar?  The principal such advantage identified by our analysis accrues in bistatic configurations wherein, for operational reasons, the transmitter must be located in a region of weak turbulence but the receiver necessarily is in a strongly-turbulent region.  Beyond that, however, there are some technological possibilities.  The ghost imager only requires a single-pixel detector, whereas the laser radar needs a detector array.  For wavelength regions in which high-performance  single-pixel detectors are available but similar-quality detector arrays are not, ghost imagers would provide active-imaging capability that laser radars could not.  A related technological advantage arises for 
 ghost imaging in multi-static configurations, in which a network of simple, small, single-pixel detectors view a target region that is floodlit by a single, large-aperture, structured-illumination transmitter.  Individual images  could be formed from each detector's outputs to capture multiple views of the target, and allow for more averaging of the target-induced speckle.  A corresponding multi-static laser radar would require high-resolution CCDs at each receiver location, making it more complicated and more expensive than the ghost imager.  
 
 \begin{acknowledgments}
This work was supported by the DARPA Information in a Photon Program under U.S. Army Research Office Grant No.\ W911NF-10-1-0404.
\end{acknowledgments}


\begin{thebibliography}{4}
\bibitem{footnote0} Ghost imagers intrinsically require multiple pulses to form an image, whereas a laser radar can provide an image from a single pulse.  For now we will allow them to use the same number of pulses, but we will return to this distinction, in Sec.~\ref{sec:Comp}, after we have derived the two systems' spatial resolution and SNR behaviors.
\bibitem{footnote2}A typical SLM would have square pixels of finite dimension and, for computational ghost imaging, be used to impress an array of random phases (uniformly distributed on $[0, 2\pi]$) on an incident laser pulse.  For analytical convenience, we are passing to a continuum limit for transverse coordinates, rather than discretizing space into pixels in these correlation functions.  Thus $2a_0$ and $\rho_0$ roughly correspond to the SLM's diameter and pixel size.
\bibitem{footnote3}The choice of a Gaussian pupil function is made for analytical convenience; a unity-transmission circular pupil could be used instead at the cost of more complicated formulas for noise terms in the SNR analysis to follow, as well as in the performance comparison with the laser radar.
\bibitem{footnote4}With this assumption, $T_p'$ appearing in our spatial-resolution formula reduces to $T_p$.
\bibitem{footnote5}Because the pulse-repetition interval, $T_p$, is fixed, letting $N\rightarrow\infty
$ will necessarily drive the time-averaged turbulence factor $\gamma$ to zero, hence it does not appear in the saturation SNR expression.
\end{thebibliography}
\end{document}